# A figure of merit-based electro-optic Mach-Zehnder modulator link penalty estimate protocol

D. M. Gill[a,*], W. M. J. Green[a], S. Assefa[a], J. C. Rosenberg[a], T. Barwicz[a], S. M. Shank[b], H. Pan[a], Y. A. Vlasov[a]

[a]*IBM T. J. Watson Research Center, 1101 Kitchawan Road, Yorktown Heights, New York 10598, USA*
[b]*IBM Systems & Technology Group, Microelectronics Division, 1000 River St., Essex Junction, Vermont 05452, USA*
*\*Corresponding author: dmgill@us.ibm.com*

We derive equations that quantify silicon Mach-Zehnder Interferometer (MZI) modulator impact upon optical link budget for NRZ transmissions based solely upon modulator extinction ratio (ER), the efficiency-loss figure of merit (FOM), and peak-to-peak drive voltage (Vpp). Our modulator link penalty equations transform the modulator efficiency-loss FOM from a simple device quality metric into a means of predicting how design and technology choices impact system margin. Our results indicate that, with a 17.8 V-cm FOM and 1 Vpp drive, designing an MZI to have an ER anywhere within the large range from 3.5-10 dB leads to nearly constant link margins, identical to within 0.5 dB.

**1. Introduction.** Complimentary metal-oxide-semiconductor (CMOS) compatible photonics technology holds promise for broadening the application space addressed by large scale CMOS manufacturing. The introduction of electro-optic processes into the CMOS manufacturing platform enables longer reach data transport, which in conjunction with the sophistication of CMOS electronics creates potential for new component, subsystem, and data transport link architectures able to address applications including advanced computing systems, data-centers, and telecommunication links. While longer reach transmission links using silicon electro-optic modulators are starting to be investigated [1], highly integrated transceivers for shorter reach data links have been commercially demonstrated [2-3]. CMOS compatible manufacture is of particular interest due to its historical reliability, low-cost high-volume manufacturing, and extensive infrastructure.

The silicon electro-optic modulator is a critical transceiver component [4-18]. The ability to monolithically integrate electro-optic modulators into larger transceivers, combined with the relatively high optical loss of plasma dispersion phase modulators, fundamentally changes the factors most relevant for optimizing these components and systems. Traditional challenges for optimizing stand-alone electro-optic MZI modulator designs, like those made in LiNbO$_3$, predominantly rely on V$\pi$, bandwidth, and impedance matching [19]. To first order, optical loss in these systems is dominated by fiber input/output (IO) connections rather than intrinsic waveguide loss. For highly integrated CMOS compatible systems, which commonly use plasma dispersion effect phase modulators [20], the optical absorption from waveguide doping in the MZI modulator RF/optic interaction region is a critical aspect of link optimization. The waveguide doping is also intrinsically coupled to V$\pi$, as well as velocity and impedance matching in traveling wave MZIs [4, 8,13,14]. Furthermore, CMOS circuitry has a maximum available drive voltage that is related to the manufacturing technology node used, which creates a fundamental trade-off between optical loss and attainable modulator extinction ratio. In concert, these issues necessitate a seminal shift in the way system metrics are balanced within plasma-dispersion-based transmitter links. In this paper we utilize the plasma dispersion phase modulator efficiency-loss FOM [4] in derived transmitter link penalty equations that help system designers understand the impact a phase modulator junction and CMOS node technology will have upon link budget. The transmitter link penalty is defined to quantify the impact a non-return to zero (NRZ) MZI transmitter will have on system performance. Analysis of these relationships reveals that the transmitter link penalty can be insensitive to large variations in the modulator extinction ratio, a surprising insight unique to silicon plasma-dispersion-based MZI transmitters.

**2. Simulation of MZI Loss and ER from V$\pi$*L and Loss in Phase Modulator.** The bias-dependent electro-optic sensitivity (denoted by the V$\pi$*L product) and loss from a plasma-dispersion-based phase modulator was investigated. Changes in phase modulator optical path length and loss were determined in a phase modulator with an interdigitated PN junction described elsewhere [7], except that the modulator presented here has doping concentrations within the electro-optic PN junction that were 1/3 of those in [7]. The resulting DC bias-dependent phase modulator $V\pi*L$ and loss are shown in Fig. 1(a).

Simulations of MZI modulator optical loss (defined as the optical loss from the MZI caused by the doped optical waveguides in the RF/optical interaction region) and DC extinction ratio (ER) versus MZI length were performed assuming a -0.5 V DC bias and 1Vpp push-pull drive, and applying the measured $V\pi*L$ and loss results to the MZI transfer function shown in Eq. 1, assuming 50/50 input/output couplers and that the MZI is biased at its quadrature point [19],

$$E = E_0\{e^{-\sigma_1 L}\exp[i\Delta\phi_1(t)] + e^{-\sigma_2 L}\exp[i\Delta\phi_2(t)] + \pi/2\}/2 \quad (1)$$

Here, $e^{-\sigma_1 L}$ and $e^{-\sigma_2 L}$ account for the optical propagation loss in each MZI arm, and $\exp[i\Delta\phi_1(t)]$ and $\exp[i\Delta\phi_2(t)]$ are time dependent optical phase shifts within the MZI in the absence of reflections. If we target a 6 dB extinction ratio (ER) we obtain a push-pull phase modulator length of ~0.33 cm and 3.68 dB of optical loss. We note that the phase modulator length represents a single RF/optical interaction length in a push-pull modulation configuration (i.e. the length of one MZI arm).

**3. Figure of Merit Definition.** We now compare the 3.68 dB loss, 6 dB ER, obtained with -1Vpp drive through the simulations in the previous section, to similar quantities derived using the modulator efficiency-loss FOM via the very simple loss calculation formula presented below. The use of a single FOM, with units of V-dB, provides significantly more insight into MZI modulator impact on link budget than previous commonly used FOMs. Since various CMOS technologies and driver design approaches supply different maximum voltage swings, the use of an efficiency-loss FOM with V-dB units allows one to accurately and quickly calculate expected optical loss once the available drive voltage and desired ER are known. The use of a V-dB FOM removes device length from explicit consideration, and simplifies the modulator system design parameter space to only ER and loss. Simplifying the design parameter space down to only ER and loss is useful because an NRZ transmitter eye closure penalty can be directly derived if the modulator ER is known, and the sum (in dB) of modulation ER penalty and transmitter loss succinctly quantifies transmitter impact upon link budget, in the absence of significant transmitter bandwidth limitations.

For example, Fig. 1b shows a plot of FOM versus Vpp bias change for our interdigitated design, where the -1V value is the $V\pi*L$ product for a -1V to 0V bias change multiplied by the optical propagation loss at -0.5 V. The loss at -0.5 V is used since the overall MZI loss is best given by the loss in one MZI arm while it is biased at its modulation quadrature point. This is true even when one MZI arm has a 0V bias and the other -1V bias on it. Fig. 1b shows the 1 Vpp FOM is 17.84 V-dB, for a 0 to -1 V bias change. If a push-pull design is used only a $\pi/2$ phase shift is required in each MZI arm. Furthermore, if we only require a 6 dB extinction ratio, then only a $0.41*\pi/2$ phase shift is required in each push-pull MZI arm when biased at the quadrature point. The push-pull MZI loss = (FOM/Vpp)*(fraction of $\pi$ phase shift required), so for a 6 dB ER MZI loss=((17.84 V-dB/1Vpp)*(0.41/2)) = 3.66 dB. This is in excellent agreement with the 3.68 dB value calculated using a careful simulation with a full MZI model and measured $V\pi*L$ and loss.

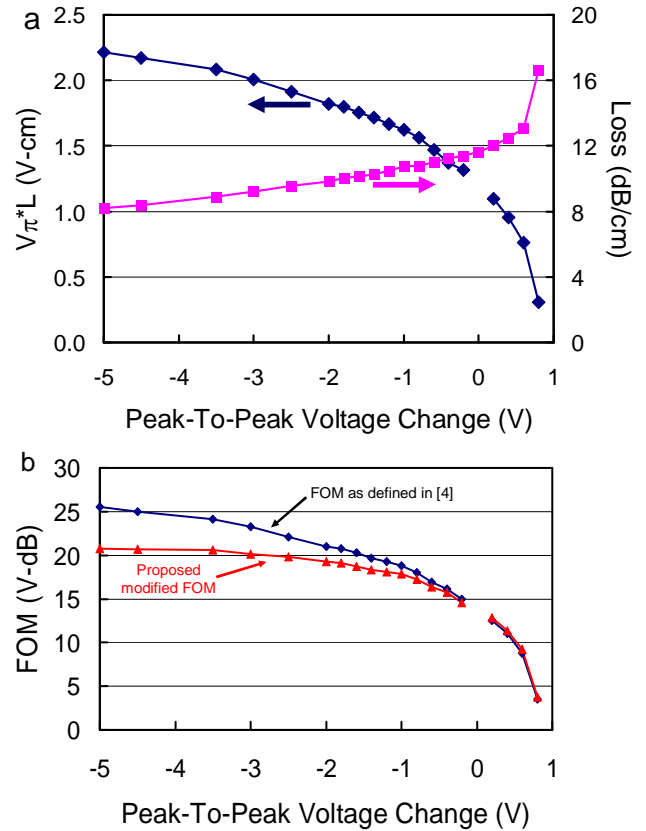

Fig. 1. Figures of merit for interdigitated PN junction plasma dispersion phase modulator a) Phase modulator $V\pi*L$ (V-cm) versus DC bias change (diamonds) and loss (dB/cm) versus DC bias (squares). b) Proposed phase modulator FOM (V-dB) versus DC bias change (triangles) and FOM as defined in [4] (diamonds).

**4. Transmitter Link Penalty Equations**. In Table 1, we use the efficiency-loss FOM to compare all modulators that we are aware of with $V\pi*L \leq 2$ V-cm for which a propagation loss is provided. Results are ranked from smallest FOM (for minimal MZI link penalty) to largest.

Once the available Vpp drive and desired ER is known, the efficiency-loss FOM can be used to easily calculate transmitter link penalty (TLP) as defined below and shown in Eq. 3 and Fig. 2. We note that Equation 3 is made up of two terms. The first term is the link penalty from a non-ideal ER, here called the ER penalty = $10*Log_{10}[(10^{ER/10}-1)/(10^{ER/10}+1)]$. The second term is the modulator optical loss and is derived from the equation loss = (FOM/Vpp)*(fraction of $\pi$ phase shift required). The 'fraction of $\pi$ phase shift required' term is calculated as a function of MZI ER assuming it is biased at its quadrature point. The derivation of the second term is simplified by noting that the (MZI maximum output) = [1-(MZI minimum output)] when biased at its quadrature point. For simplicity, our analysis considers only DC operation of the MZI. We note that incorporating link penalties from transmitter bandwidth limitations is highly dependent upon the choice of fabrication technology, and is beyond the scope of this work.

**Table 1. Comparison of published phase modulator designs using the efficiency-loss FOM defined in [4]. This FOM definition is used since the 0V loss is most commonly published.**

| Institution | Pub date | Loss (dB/cm) | $V\pi*L$ (V-cm) | FOM (V-dB) | Ref |
|---|---|---|---|---|---|
| IBM - Low doped | Presented | 11.5 | 1.6 | 18.4 | - |
| CAS | July-12 | >10.0* | 1.5 | < 20 | 5 |
| CAS | Jan-12 | 16.7 | 1.2 | 20.7 | 6 |
| IBM - High doped | Nov-12 | 31.8 | 0.8 | 25.4 | 7 |
| Bell Labs-IME | Feb-12 | 15.0 | 1.9 | 27.9 | 8 |
| Sandia | Feb-10 | 31.0 | 1.0 | 31.0 | 9 |
| Astar | Mar-10 | 70.0 | 0.5 | 35.0 | 10 |
| ALU-BAE | Jan-09 | 18.5 | 2.0 | 37.0 | 11 |
| ETRI-Korea | Dec-11 | 32.0 | 1.6 | 51.0 | 12 |
| CAS | Mar-12 | 62.5 | 1.3 | 80.0 | 13 |

*Note that [5] only gives the excess loss created by the waveguide doping, which precludes a precise FOM determination.

TLP = {ER penalty in dB} + {modulator loss in dB}

$$TLP = \left\{10Log_{10}\frac{(10^{ER/10}-1)}{(10^{ER/10}+1)}\right\} + \left\{\frac{FOM}{2Vpp}\left(1-\frac{4\arccos\left(\sqrt{\frac{1}{1+10^{\frac{ER}{10}}}}\right)}{\pi}\right)\right\} \quad (3)$$

We note that even though the modulator loss term in Eq. 3 was derived from an idealized MZI transfer function, with no differential loss between MZI arms, the calculated ER is still within a few percent of results obtained with differential loss included. We also reiterate that the penalty calculated from Eq. 3 is for a push-pull modulator, since the term (FOM/2Vpp) has the factor of 2 in its denominator. From Eq. 3 we calculate that a 6 dB NRZ ER gives 2.2 dB of modulation ER penalty, which added to the 3.65 dB of modulator loss results in 5.85 dB of TLP. Equation 3 also gives insight into trade-offs between available drive voltage and link penalty, as shown in Fig. 2. Figure 2 shows that with 1 Vpp drive (0 to -1 V bias change) and a FOM=17.8 V-dB, a wide range of designs offer minimal link penalty. There is only ~0.5 dB change in penalty for ERs ranging from 3.5 dB to 10 dB. This behavior exemplifies how designing a longer MZM to increase extinction ratio will also increase optical loss, which in this case makes the output optical modulation amplitude (OMA) nearly constant for a 3.5 dB to a 10 dB ER. The insensitivity of TLP to ER (near the minimum TLP point for a given drive voltage) is a unique characteristic of plasma-dispersion-based silicon MZI modulators, a surprising insight that must be accounted for in optimizing an optical link.

Figure 2 also shows that 2 Vpp drive (0 to -2 V bias change) reduces the minimum link penalty by ~2dB, while 0.6 Vpp drive increases the minimum penalty by ~1.5 dB. Once again, this analysis assumes transmitter bandwidth does not limit performance. Any link penalty associated with bandwidth limitation must be added separately once a specific transmitter driver design is identified.

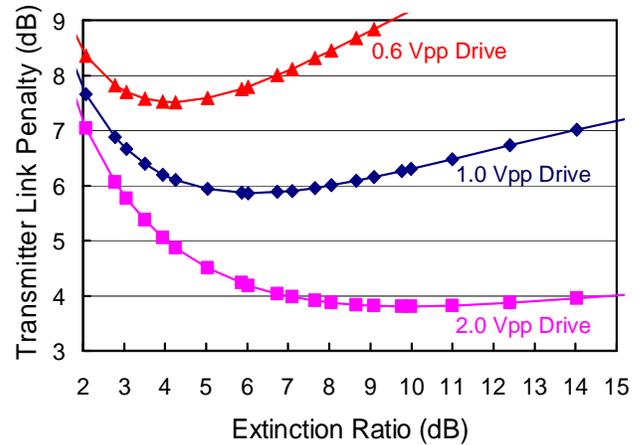

Fig. 2. Transmitter link penalty (NRZ DC extinction-ratio-based modulation penalty + optical loss) for FOM=17.8 V-dB calculated from Eq. 3 for a 0.6 Vpp (triangles), 1.0 Vpp (diamonds), and 2 Vpp drive (squares) versus ER.

The derivative of Eq. 3 was set equal to zero, and solved for ER, giving the ER required (in dB) to obtain the minimum TLP as a function of FOM and Vpp, Eq. 4. Equation 4 was used to determine the ER required for a minimum TLP versus FOM for a push-pull drive voltage of 1Vpp. This result was then used in Eq. 3 to derive the minimum TLP as a function of FOM. The values of minimum TLP and ER at minimum TLP are plotted as a function of FOM in Fig. 3. For example, the figure illustrates that for a 1Vpp drive, a FOM of 15 V-dB gives a minimum TLP of ~5.5 dB, while a FOM of 30 V-dB results in ~8 dB minimum TLP. This simple example shows how Eq. 3 and Eq. 4 can serve as valuable tools to rapidly identify the physical phase modulator device-level FOM and CMOS system-level design requirements needed to meet link budget targets.

$$ER_{MinLoss} = 20\log_{10}\left\{\frac{-1+\sqrt{1-4\left(\frac{\ln 10}{20\pi}\right)^2\left(\frac{FOM}{Vpp}\right)^2}}{2\left(\frac{\ln 10}{20\pi}\right)\left(\frac{FOM}{Vpp}\right)}\right\} \quad (4)$$

**5. Conclusion.** We have presented straightforward equations to calculate MZI transmitter contributions to NRZ link penalty as a function of MZI ER, CMOS Vpp drive, and modulator efficiency-loss FOM, which allows system designers to easily compare different phase modulator junctions and CMOS node technologies in a quantified way. Our analysis surprisingly indicates that if modulator bandwidth does not limit performance, there is ~0.5 dB change in transmitter link penalty if the MZI is designed to have an extinction ratio ranging from 3.5 dB to 10 dB, assuming a 1 Vpp drive and a 17.8 V-cm FOM. The analytic relations presented serve as valuable tools for designers to rapidly identify optimal MZI design targets and device-level modulator FOM requirements for system-level link budget targets.

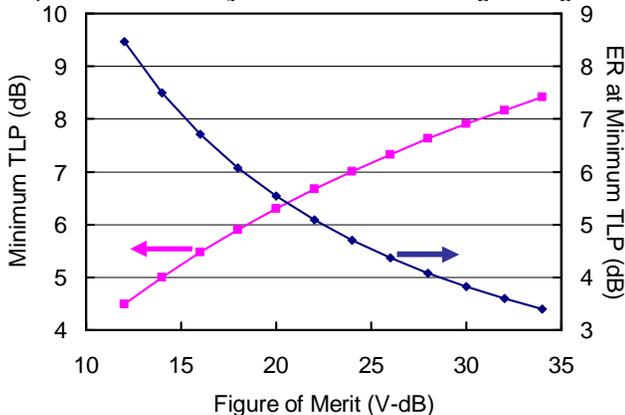

Fig. 3. Minimum transmitter link penalty (squares) and the corresponding MZI ERs (diamonds) Vs FOM for a 1 Vpp drive.

**5. Acknowledgements.** We gratefully acknowledge Jonathan Proesel for helpful discussions and insights that contributed to this paper.